\documentstyle[12pt,epsf]{article}
\begin{document}

\newcommand{\sheptitle}
{Leptogenesis in a Realistic Supersymmetric Model
of Inflation with a Low Reheat Temperature}

\newcommand{\shepauthor}
{M. Bastero-Gil$\dagger$ and S. F. King$^\ast$ }

\newcommand{\shepaddress}
{$\dagger$ Scuola Normale Superiore, Piazza dei Cavalieri 7, \\
 56126 Pisa, Italy \\
$^\ast$Department of Physics and Astronomy,
University of Southampton, \\ Southampton, SO17 1BJ, U.K.}

\newcommand{\shepabstract}
{We discuss leptogenesis in a realistic supersymmetric 
model of inflation with a low reheat temperature 1-10 GeV.
The lepton asymmetry is generated by a decaying right handed sneutrino, 
which is produced after inflation during preheating. 
The inflationary model is based on a simple variant of the
Next-to-Minimal Supersymmetric Standard model (NMSSM) which 
solves the $\mu$ problem,  
called $\phi$NMSSM, where the additional singlet $\phi$
plays the role of the inflaton in hybrid (or inverted hybrid) type models. 
The model is invariant under an approximate
Peccei-Quinn symmetry which also solves the
strong CP problem, and leads to an invisible axion with 
interesting cosmological consequences.
We show how the baryon number of the universe and the nature of cold
dark matter are determined by the same parameters controlling the
strong CP problem, the $\mu$ problem and the neutrino masses
and mixing angles.}

\begin{titlepage}
\begin{flushright}
SNS-PH/00-17 \\
hep-ph/0011385\\
\end{flushright}
\vspace{.1in}
\begin{center}
{\Large{\bf \sheptitle}}
\bigskip \\ \shepauthor \\ \mbox{} \\ {\it \shepaddress} \\ \vspace{.5in}
{\bf Abstract} \bigskip \end{center} \setcounter{page}{0}
\shepabstract
\begin{flushleft}
\today
\end{flushleft}
\end{titlepage}

\section{Introduction} 

Leptogenesis is an attractive mechanism which has been proposed 
to generate the
observed baryon asymmetry of the Universe (BAU) \cite{yanagida1,
luty}. The mechanism
involves the out-of-equilibrium decay\footnote{Models of leptogenesis
based instead in the cosmological evolution of flat directions which carry
lepton number can be found for example in
Ref. \cite{campbell,others}.}  of a heavy right handed neutrino 
$N_R$ (or sneutrino $\tilde{N}_R$ \cite{yanagida2, campbell}). The net
lepton number L produced in the decay is then  
reprocessed into baryon number B by anomalous (B+L) violating
sphaleron interactions, which otherwise conserve (B-L) \cite{kuzmin}. 
The same physics that
allows the right handed neutrinos to decay into light leptons
is also responsible for
a see-saw neutrino mass matrix. Combining the see-saw
mechanism with the latest experimental data on neutrino masses
\cite{superK} seems 
to favour a scale for the  right handed neutrino mass $M_R$ in the
range $10^{7}$ - $10^{14}$ GeV. 

What mainly distinguishes the different scenarios of leptogenesis that can be
found in the literature is the production mechanism for the heavy
 $N_R$  (or $\tilde{N}_R$).  
The lepton (baryon) asymmetry has to be produced at some early
stage in the cosmological evolution of the Universe, at some point
after inflation ends and before the time of nucleosynthesis. 
In a sense, leptogenesis/baryogenesis is closely related to the
inflationary dynamics and the post-inflationary reheating era.
The right handed sneutrino could be itself the inflaton
\cite{yanagida2}, with L 
generated during the reheating period. If this is not the case, 
and the reheating temperature ($T_{RH}$) after inflation is larger than
$M_R$, the 
heavy (s)neutrinos can be thermally produced after reheating, and the
final lepton  asymmetry will depend on the out-of-equilibrium
conditions at the time they decay \cite{yanagida1,luty,lepto}. 
However, in supersymmetric models
such a large $T_{RH}$ may be in conflict with the standard bound 
$T_{RH} < 10^9$ GeV in order to avoid an overabundance of gravitinos
\cite{gravitinos}.  
On the other hand, if $T_{RH} < M_R$  the right handed
(s)neutrinos would have to be 
produced\footnote{Strictly speaking, the condition $T_{RH} < M_R$
does not rule out 
thermal production of heavy neutrinos. In most of inflationary models
reheating is not instantaneous, and the maximum temperature $T^{max}$
reached is 
usually much larger than $T_{RH}$ \cite{kolb}, and we could have
$T_{RH} < M_R < 
T^{max}$. Like in GUT baryogenesis, there could be models where the
decay of the inflaton into heavy neutrinos may be suppressed or
forbidden, but still they could be thermally produced during the long
period of reheating \cite{tmax1}.}  
by the out-of-equilibrium inflaton
decay, either in perturbative decays \cite{yanagida3} or by parametric
resonance \cite{peloso} during 
preheating. In this way, the out-of-equilibrium condition for
baryogenesis/leptogenesis is automatically satisfied. The other two
requirements, baryon number violation and C and CP violation, will be
provided  by the sphaleron  interactions and complex
phases in the neutrino Yukawa couplings respectively. 

In this paper we shall extend the model for inflation proposed in
Ref. \cite{us} in order to include neutrino masses and
implement leptogenesis. 
The inflation model is based on the next-to-minimal supersymmetric
model and provides an intermediate scale solution to the $\mu$
problem, and the strong CP problem via the Peccei-Quinn mechanism \cite{us}.
The supergravity version of the model \cite{sugra} solves the $\eta$
problem via the implementation of a no-scale mechanism,
provides F-term inflation from the moduli fields which are 
stabilised before and after inflation and there is no moduli problem
or gravitino problem. 
The inflationary model is of the hybrid type, characterised by a not too
large scale for the vacuum energy, $V(0)^{1/4} \simeq 10^8\, GeV$  and
a very low reheating temperature $O(1\,GeV)$. It is interesting to
study leptogenesis within models with such a low reheat temperature
since in such models 
thermal production of Majorana neutrinos during or after reheating
is impossible, so
the production mechanism will rely on preheating the fields 
which occurs during the
oscillatory period following the end of inflation. 
This model is particularly interesting since the oscillating inflaton
fields at the end of inflation do not couple directly to the
sneutrinos, but only indirectly via a coupling to the Higgs doublets.
Thus the production of sneutrinos during preheating is linked
also to the production of Higgs scalars, and since the Higgs scalars
decay into both radiation and neutralinos it becomes possible to 
relate the relic density of the lightest neutralinos to the
baryon number of the universe. Relativistic axions are also produced
during reheating but these are red-shifted away, although later on 
non-relativistic axions are additionally produced by the usual
misalignment mechanism and these will contribute to cold dark matter.

The main advantage of studying a realistic 
supersymmetric particle physics model of
inflation is that questions such as the nature
of cold dark matter and baryogenesis via leptogenesis
are related, and determined by
the same parameters which control the particle physics questions of
the $\mu$ problem, the strong CP problem, and neutrino masses
although as we shall see
there are many uncertainties at present and many of our estimates
will have errors of one or two orders of magnitude.

The layout of the rest of the paper is as follows.
In section 2 we
summarise the main properties of the model, and introduce the right
handed neutrinos with the usual superpotential suitable for neutrino
physics and leptogenesis. An estimation of the
lepton asymmetry produced is given in section 3. Because the right
handed neutrinos will decay  in less than a
Hubble time, much before the inflaton has time to decay, we need to
check how much of the asymmetry survive the reheating era. This is
done in section 4. In section 5 we present our conclusions. 

\section{A model for inflation and leptogenesis.} 

The model of inflation we have proposed is based on the superpotential
\cite{us}: 
\begin{equation}
W= \lambda N H_1 H_2 + \kappa \phi N^2 \,,
\label{phinmssm}
\end{equation}
where $H_1,\,H_2$ are the Higgs doublets and $\phi$, $N$ are gauge
singlets. The superpotential is invariant under a $U(1)_{PQ}$
Peccei-Quinn symmetry, which is broken during and after inflation by
the vevs of $\phi$ and $N$.
The vacuum energy $V(0)$, needed during
inflation, originates from an F-term of the effective sugra theory
\cite{sugra}. 
The Higgs doublets play no role during inflation, $\phi$ is the
inflaton, and $N$ is the second singlet needed to end hybrid inflation.  
Imposing the slow-rolling and COBE constraints for inflation gives the
order of magnitude results:
\begin{equation}
\kappa\sim 10^{-10}\,,\; \langle \phi \rangle \sim \langle N \rangle
\sim 10^{13}\, GeV,\,\; V(0)^{1/4} \sim 10^{8}\, GeV \, \; H(0) \sim
O(MeV) \,, 
\label{v0}
\end{equation} 
with $H(0)$ the Hubble parameter during inflation. In order to have an
effective $\mu$ term in the Higgs sector of 
the correct order of magnitude, we require $\lambda \sim \kappa$.
The smallness of the couplings is accounted for by the use of
higher dimensional operators, so the superpotential in Eq.\ref{phinmssm}
should be viewed as an effective superpotential which originates
from some intermediate scale solution to the $\mu$ problem,
as discussed elsewhere \cite{us}.
Due to the smallness of the (effective) couplings, the fields $\phi$ and
$N$ are very long-lived, decaying mainly into axions with a decay rate 
$\Gamma_\phi \sim 10^{-17}\,GeV$. Considering only the standard
perturbative reheating period following inflation, this will give rise
to a reheating temperature of order a few GeVs, much below the
electroweak scale.

Right-handed majorana neutrinos are introduced in the model with the
usual superpotential:
\begin{equation}
W= M_{R_i} N_{R_i} N_{R_i} + \lambda_{{L R}_{ij}} L_i H_2
N_{R_j} \,,
\end{equation}
written in the eigenstate basis for the $N_{R_i}$. 
The right handed neutrino masses and Yukawa couplings $\lambda_{LR}$
has to be such that they reproduced the observed properties of the
light neutrino spectrum. The recent data from Super-Kamiokande
supports $\nu_\mu \leftrightarrow \nu_\tau$ oscillations with $\sin^2 
2 \theta_{23} > 0.88$ and a mass squared splitting $\Delta m^2_{23}
\simeq (1.5-5) \times 10^{-3}\, eV^2$. It also favours the large
mixing angle solution for solar neutrino mixing, with $\sin^2 2
\theta_{12} \sim 0.75$ and $\Delta m^2_{12} \sim 2.5 \times 10^{-5} \,
eV^2$, although others solutions are not excluded.    
For numerical estimations, and as a working example, we will use the
results of a recent estimate of all quark and lepton masses and
mixing angles based on a string-inspired Pati-Salam model \cite{steve1},
although we shall only be concerned with the leptonic part of this model.
Due to the gauged $SU(2)_R$ symmetry the model predicts three
right-handed neutrinos, and the heaviest one is the one associated
with the third family with a mass of $10^{14}$ GeV. Although this
is the heaviest it nevertheless plays the dominant role
in generating the atmospheric masses and mixing angle, due to the
Yukawa structure of the model, leading to an automatic neutrino mass
hierarchy, according to the
single right handed neutrino dominance mechanism \cite{steve},
and bi-maximal mixing. In terms of the
heaviest right handed scale of the order $M_{R_3}
\simeq 10^{14}\,GeV$, and the Wolfenstein expansion parameter
$\lambda\simeq 0.22$, the right handed neutrino masses and couplings are
given by: 
\begin{eqnarray}
 M_{R_i} &\sim& 10^{14} \left( \lambda^{9}, \,\lambda^{5}, \,1 \right)
         \,GeV 
         \sim \left( 10^{8},\, 10^{10},\, 10^{14}\right) \, GeV\,, 
\label{masses} \\   
\lambda_{LR} &\sim& \left( \matrix{  \lambda^{8} & \lambda^{4} &
\lambda^2 
\cr \lambda^7 & \lambda^4 & \lambda
\cr 
\lambda^7 & \lambda^4 & 1
\cr} \right) \label{couplings} \,, 
\end{eqnarray}
Note that the lightest right-handed neutrino of mass 
$10^{8}\,GeV$ is significantly lighter
than $10^{14}\,GeV$ but does not give the dominant contribution to
physical neutrino masses due to its suppressed Yukawa couplings,
although it is light enough to be produced through preheating.

Given  that in the inflationary model  $T_{RH} \ll V(0)^{1/4} <
M_{R_i}$, the $N_{R_i}$ 
($\tilde{N}_{R_i}$) fields cannot be produced  
thermally at any stage, nor in the perturbative decay of $\phi$ and
$N$ fields. Therefore, we depend upon preheating for that.

Due to the lack in the superpotential of a direct Yukawa coupling
between the singlets superfields $\phi$, $N$ and  $N_R$,  
preheating of the right handed neutrinos does not look
possible. The situation is different for the scalar components,
because of the coupling $\lambda_{LR}$ with the doublet Higgs $H_2$
and the leptons $L_i$. This will induce a term in the scalar potential of the
form: 
\begin{equation}
V= \cdots + |\lambda_{LR}|^2_{kj} |H_2|^2 \tilde{N}_{R_j}
\tilde{N}^\ast_{R_k}+ \cdots \,.
\end{equation}
Large oscillations in the Higgs fields will be induced through their coupling
to the singlets in Eq.\ref{phinmssm}, 
and in turn we expect this to trigger the preheating
of the sneutrinos. The lepton asymmetry will
be generated by the decay of the sneutrinos instead of that of the
neutrinos. We notice also that the maximum possible $T$ that we
can reach during reheating is going to be smaller than the $M_{R_i}$
masses (at most $T^{max} \sim 10^8 \, GeV \sim M_{R_1}$), so once
produced, the  sneutrinos will remain out of equilibrium and will
decay faster than say the inflaton.  

\section{Lepton/Baryon asymmetry}

Preheating of right-handed sneutrinos provides the seed we need 
for leptogenesis, i.e., a non-zero number density of the order
\begin{equation} 
n_{\tilde{N}_{R_i}} \sim c_i \frac{V(0)}{M_{R_i}} \,,
\end{equation}
where $c_i$ parametrize the fraction of the total vacuum energy which
is transfered to the sneutrinos during preheating. 
Given the hierarchy in masses, it is not
unreasonable to assume $c_3 \ll c_2,\, c_1$, whilst we will take 
 $c_2 \sim c_1 \sim c \sim O(1-0.1)$.
CP violation in the decay of $\tilde{N}_{R_i}$ comes from the interference
between the tree-level and one-loop amplitudes
\cite{luty,lepto,covi,epsilon}. The CP asymmetries given by the
interference with the one-loop vertex amplitude are \cite{luty,lepto}:
\begin{eqnarray}
\epsilon_i &=& \frac{\Gamma ( \tilde{N}_{R_i} \rightarrow \tilde{l} +
    H_2 ) - \Gamma(\tilde{N}^\dagger_{R_i} \rightarrow
    \tilde{l}^\dagger + H^\dagger_2)}
{\Gamma( \tilde{N}_{R_i} \rightarrow \tilde{l} +
    H_2) + \Gamma(\tilde{N}^\dagger_{R_i} \rightarrow
    \tilde{l}^\dagger + H^\dagger_2) }
\nonumber \\
    &=&\frac{1}{8\pi(\lambda^\dagger_{LR} \lambda_{LR})_{11}}
\sum_j \left( Im \left[ (\lambda^\dagger_{LR}
\lambda_{LR})_{1j}\right]^2 \right) f(M_{R_j}^2/M_{R_i}^2)\,,
\end{eqnarray}   
where
\begin{equation}
f(x) = \sqrt{x} \left[ 1 - (1+x) \log \left( \frac{1+x}{x} \right)
\right] \,.
\end{equation}
The interference with the absorptive part of the one-loop self-energy 
also gives a contribution to the asymmetry, which in general is the same
order as those given above, unless the (s)neutrinos were 
almost degenerate in which case could it be much larger \cite{covi,
epsilon}.  

As an example in order to estimate the values of $\epsilon_i$, we will
consider the model given in Eqs. \ref{masses} and  
\ref{couplings}.  Assuming maximal CP violation ( $Im[\cdots]^2
\sim | \cdots|^2$), we can see that the asymmetries will be dominated
by the larger couplings to the third generation of leptons, with:
\begin{eqnarray} 
\epsilon_1 & \sim & \frac{ | \lambda_{LR} |^2_{33}} {8 \pi}
\frac{M_{R_1}}{M_{R_3}} \sim \frac{ \lambda^9 }{8 \pi} \sim 10^{-7} - 10^{-8}\,, \\
\epsilon_2 & \sim & \frac{ | \lambda_{LR} |^2_{33}} {8 \pi}
\frac{M_{R_2}}{M_{R_3}} \sim \frac{ \lambda^4 }{8 \pi} \sim 10^{-4} - 10^{-5}\,, \\
\epsilon_3 & \sim & \frac{ | \lambda_{L R} |^2_{32}} {8 \pi}
\frac{M_{R_2}}{M_{R_3}} \sim \frac{ \lambda^{13} }{8 \pi} \sim
10^{-10}- 10^{-11} \,.
\end{eqnarray}
We remark again that these are only order of magnitudes estimations,
with large uncertainties in their values. Besides, the values of
$\epsilon_i$ are model dependent. Other texture models with values of
the Yukawas conistent with the experimental data on neutrinos, and
similar hierarchy among the right handed neutrino masses, could give
rise to a larger asymmetry such as $\epsilon_1 \sim 10^{-6}$
\cite{example}.  

The decay of the sneutrinos occurs fast enough to neglect any effect
due to the expansion of the Universe, and the lepton asymmetry is then
given by: 
\begin{equation}
n_{B-L} \simeq \epsilon_i n_{\tilde{N}_{R_i}} \approx \epsilon_i
 c_i \frac{V(0)}{M_{R_i}} \,.
\end{equation} 
The decay of the heaviest right handed sneutrino (if produced) will 
give rise to a negligible lepton asymmetry, whilst that generated in
the decay of $\tilde{N}_{R_1}$ and $\tilde{N}_{R_2}$ are comparable because
$\epsilon_1/ M_{R_1} \sim \epsilon_2/ M_{R_2} \sim 1/M_{R_3}$.  
This is then converted into baryon number by B+L violating sphalerons
interactions (which are in equilibrium for temperatures in the
interval $\sim [200, 10^{12}]$ GeV) \cite{nbnbl}, 
\begin{equation} 
n_B = -\frac{8}{23} n_{B-L} \,,
\end{equation}  
and finally at the time of nucleosynthesis we will have:
\begin{equation}
\left. \frac{n_B}{s} \right |_{nucl.} \approx \frac{8}{23} \gamma
\epsilon_i \frac{( c 
V(0))^{1/4}}{ M_{R_i}} \sim \frac{\gamma}{23 \pi} |\lambda_{L
R}|^2_{33} \frac{( c V(0))^{1/4}}{ M_{R_3}} \,, 
\end{equation}
where we have used $s=(2 \pi^2/45) g_* T^3$ evaluated at the time the
leptons are produced, i.e., $T\simeq .3 (c V(0))^{1/4}$, with the effective
number of relativistic degrees of freedom $g_* \sim 100$. The factor
$\gamma$ accounts for the dilution due to possible 
entropy production during reheating. 
Substituting the values of $V(0)$ and $M_{R_3}$
Eqs. (\ref{v0}) and (\ref{masses}), with $\lambda_{{LR}_{33}} \simeq
1$, we obtain: 
\begin{equation}
\left. \frac{n_B}{s} \right |_{nucl.} \sim  c^{1/4} \times 10^{-8} \gamma \,.  
\end{equation}
In order to explain the observed baryon asymmetry $n_B/s \sim
10^{-10}$ \cite{eta} we cannot allowed much entropy  (radiation)
production during the reheating era. In the next, we will try to
estimate the factor 
$\gamma$ based on simple assumptions.

\section{Preheating/Reheating}

During the oscillations of the background fields $\phi$ and $N$,
particles can be produced by parametric resonance (preheating)
\cite{preheating1,preheating2,preheating3} 
much before the inflaton has time to decay perturbatively, being this
in general a more efficient mechanism of particle production
than standard perturbative decay.  

Even if the couplings are very small, the amplitude of the
oscillations is large enough to ``preheat''  the modes of the scalar
fields $\phi$ and $N$. Beside, the value of Hubble parameter is small
in the model, which allows for a large number of oscillations in a
Hubble time, before they start to feel the effect of the expansion. 
Because hybrid inflation ends in a phase transition, with the effective
squared mass of the $N$ field changing sign, production of $\phi$ and
$N$ quanta is very efficient during the first few oscillations of the
background fields \cite{preheatingus}. Due to the coupling $\lambda$
between the 
Higgses and the $N$ field, we also expect to preheat the Higgs fields
$H_i$ in a similar way. The 
evolution equations for the Higgs quantum fluctuations are indeed
analogous to those of the singlets, and we can assume similar number
densities for both. The sneutrino fields $\tilde{N}_{R_i}$  are
therefore preheated  
through the Higgses, with the lightest one more likely to be produced,
and they will be clearly out of equilibrium.
We may also preheat axions, fermions, etc.. but with
much smaller number densities\footnote{Due to the smalleness of the
couplings, non-thermal production of gravitinos \cite{gravitinos2} is
 not a problem \cite{BGM}.}.
     
Preheating is efficient only in producing very low frequency
modes. Nevertheless, rescattering effects will allow to excite higher
frequency modes and redistribute the energy density.   
Based on the results for the singlets \cite{preheatingus}, we may
estimate that after just 3-4 
oscillations a fraction  of the vacuum energy has been transfered to
the singlets and the other fields quanta, with more or less equal energy
densities. The typical time scale for this to happen is given by:
\begin{equation}
\Delta t_{preh} \simeq \frac{ 2 \pi \Delta N_{osc}}{M_\phi} \sim 10^{-2}
\, GeV^{-1} \,,
\end{equation}
where $\Delta N_{osc}$ counts the number of oscillations, and $M_\phi$ is the
mass of the fields $\phi$ and $N$ in the global 
minimum (and therefore the typical frequency of their oscillations):
\begin{equation}
M_\phi = \kappa \langle N \rangle = O(1\, {\rm TeV}) \,.
\end{equation}

The decay rates of the fields involved, $\tilde{N}_{R_{1,2}}$, $H_i$, and
singlets $\phi$ and $N$, can be estimated as:
\begin{eqnarray} 
\Gamma_{\tilde{N}_{R_1}} &\simeq& \frac{|\lambda_{LR}|_{11}^2}{8 \pi} M_{R_1}
\approx O(0.01) \, GeV  
\,, \\
\Gamma_{\tilde{N}_{R_2}} &\simeq& \frac{|\lambda_{LR}|_{22}^2}{8 \pi} M_{R_2}
\approx
O(10^4) \, GeV \,, \\
\Gamma_{H_i} &\simeq& \frac{g^2}{8 \pi} M_H \approx O(10)\, GeV \,, \\
\Gamma_{\phi} &\simeq& \frac{\kappa^2}{8 \pi} M_\phi \approx
O(10^{-17})\, GeV \,, 
\end{eqnarray}
where $g$ is the electroweak coupling constant, and the mass of the
Higgses are $M_{H_i} \simeq O( 100\, GeV - 1\, TeV)$. 
The sneutrinos  $\tilde{N}_{R_2}$ will tend to decay immediately after they
are produced, 
$\Gamma_{\tilde{N}_{R_2}}^{-1} \ll \Delta t_{preh}$, and its decay
products quickly thermalize by scattering from each other
\cite{ellis}, given that:
\begin{equation}
\Delta n_{l} \sigma_{sc} > H
\end{equation}
where $\Delta n_{l}$ is the number density of the light degrees of
freedom $\Delta n_{l} \simeq n_{\tilde{N}_2}$, $\sigma_{sc}
\propto M_{R_2}^{-2}$, and $H$ is the Hubble parameter of the order
$O(MeV)$.   
No back reaction is 
expected from them, except that part of the vacuum energy is converted
into radiation with a temperature $T_0 \simeq 0.3 c_2^{1/4}
V(0)^{1/4}$. 
The fields $H_i$ and $\tilde{N}_{R_1}$ do not decay before $\Delta
t_{preh}$.    
Therefore, back reaction effects due not only to the singlets but also
to the Higgses (and eventually $\tilde{N}_{R_1}$) will soon slow down and
suppressed the rate of production of particles during preheating, in
less than a Hubble time $H^{-1}$. 

At this point we can consider that the Universe has been reheated up to a
temperature $T_0 \sim O(10^8\, GeV)$, but  with a non-negligible
fraction of the energy 
still in the form of cold oscillations and singlets, Higgses and
$\tilde{N}_{R_1}$. The right handed sneutrinos $\tilde{N}_{R_1}$ also
decay out-of-equilibrium 
in a time $\delta t \simeq 
\Gamma_{\tilde{N}_{R_1}}^{-1} \ll H^{-1}$, transferring its energy to
the thermal bath.      
On the other hand, the decay rate of the Higgses will now be
suppressed by the factor $M_H/T_0$, rendering it quite inefficient.  
Therefore, after
preheating becomes inefficient, back reaction and rescattering effects
take place, and both $\tilde{N}_{R_2}$ and $\tilde{N}_{R_1}$ decay, we
are left with the vacuum 
energy distributed among the singlets, Higgses, and radiation.

Preheating can become very inefficient, but will not neccessarily
stop as long as $\phi$ and $N$ continue to oscillate (that is,
there is some energy density left in these fields), and production of
Higgses from the singlets might continue in a very narrow resonance
regime \cite{preheating1}, at the same time than standard reheating.  
On the other hand, we do not expect this effect to be enough to further
induced the production of the much heavier right handed sneutrinos. 
The preheating/reheating era will end when the singlets finally 
decay, at a time $t \simeq \Gamma_\phi^{-1}$. 

If the masses of the singlets are smaller than the
lightest Higgs mass, the fields $\phi$ and $N$ decay 
 predominantly into axions, with the branching ratio
into others particles being much smaller. The axions behave as
relativistic particles, but they do not thermalize
\cite{thermalaxions}, that is, their interaction rate always remains 
smaller than the Hubble expansion parameter.  
The axion interaction rate $\Gamma_{AI}$ is given by:
\begin{equation}
\Gamma_{AI} = \langle \sigma_a |v| \rangle n_R
\end{equation}
where $n_R$ is the radiation number density $n_R \simeq T^3/\pi^2$,
and $\sigma_a$ is the 
axion cross section for scattering off the thermalize radiation. 
On dimensional grounds, the cross section can be written as\footnote{The
coupling $\alpha_a$ is either due to tree-level interactions, and
therefore further suppressed by a factor $T^{-2}$ when the temperature is
larger than the typical mass scale of the particle exchanged, or due
to loops effects, which are suppressed by a factor $1/(8
\pi^2)^2$. Therefore, until the temperature drops near $O(1\,GeV)$ we
have $\alpha_a \sim 10^{-4}$.}  $\sigma_a=
\alpha_a/f_a^2$, 
$f_a$ being the axion decay constant $~ 10^{12} - 10^{13} \, GeV$. 

Let us define $t_0$ as the initial time after
the initial burst of radiation produced by the sneutrinos decay. At
this time, 
the Universe is at a temperature $T_0 \sim O(10^8 \, GeV)$ but still 
$H \sim V(0)^{1/2}/ \sqrt{3} M_P$. 
The ratio $\Gamma_{AI}/H$ is then: 
\begin{equation}
\left. \frac{\Gamma_{AI}}{H}\right |_{t_0} \simeq \frac{\alpha_a}{f_a^2}
\frac{T_0^3}{\pi^2} \frac{\sqrt{3} M_P}{V(0)^{1/2}} \simeq 10 \alpha_a
c^{3/4} \left(\frac{10^{12} \, GeV}{f_a}\right)^2 < 1 \label{axion1}\,.
\end{equation}   
After the Universe becomes radiation dominated at a
 temperature $T_{RD}$ we will have instead:
\begin{equation}
\left. \frac{\Gamma_{AI}}{H}\right |_{t_0} \simeq \frac{\alpha_a}{f_a^2}
\frac{1}{\pi^2} \left( \frac{90} {\pi^2 g_T} \right)^{1/2} T
M_P < 10^{-5} \alpha_a \left(\frac{10^{12} \, GeV}{f_a}\right)^2
T_{RD} < 1 \label{axion2}\,,
\label{ratio}
\end{equation}
where we have used $H \simeq (\pi^2 g_T/90) T^2/M_P$, and $g_T$ is the
effective number of relativistic degrees of freedom at $T$. 
The last inequality in Eq. (\ref{ratio}) follows because the factor
$\alpha_a$ becomes 
$O(1)$ only when $T\sim O(1\,GeV)$. 
If radiation is being produced while the Universe is matter dominated,
the number 
density and Hubble parameter evolve as $n_R \propto a^{-1/2}$ and $H
\propto a^{-3/2}$, with $a$ the scale factor, so the ratio 
$\Gamma_{AI}/H$ increases in time. In a radiation dominated Universe
they will both scale as $a^{-1/2}$. In either case, 
Eqs. (\ref{axion1}) and (\ref{axion2})  ensure that the axions never come
into equilibrium. This means that we do not
expect radiation to be produced from the singlets (inflaton), contrary
to most models of inflation. Any extra radiation will come eventually from the
Higgses decay. 

Let us now briefly summarise the above discussion. We have argued that
right-handed sneutrinos are produced 
during the initial period of preheating, but they decay rapidly into
leptons and Higgses. We are now 
interested in the evolution of the Universe from this time until the
time of reheating $t_{RH}$, defined as the time at which the singlets
completely decay\footnote{In general, the reheating time will
coincide with the time at which the Universe becomes radiation
dominated $t_{RD}$. However in the present context this  may not be
the case, see below.}. In this interval there is an interplay between
the energy density of the oscillating inflaton fields $\rho_\phi$, the
number density of Higgses $n_H$, the energy of the axion fields
$\rho_{axion}$, and the energy density in radiation $\rho_R$. We shall
model this as follows. $\rho_\phi$ will be steadily reduced due to
continual production of axions (through the standard perturbative rate
$\Gamma_{\phi a}$) and Higgses (through inefficient preheating). The
axions behave as relativistic matter, but they stay out of
equilibrium. Higgses are created and annihilate into radiation with a
thermal-averaged crossed section $\langle  \sigma_H|v| \rangle$ and
decay rate $\langle \Gamma_H \rangle$. Therefore, the radiation
density $\rho_R$ receive contributions from the Higgses but not from
the singlets. 
On the other hand, among the decay products of the Higgses we will
find also neutralinos, the lightest of them being a candidate for cold
dark matter. The standard calculation of their relic abundance
\cite{neutralinos} depends mostly on their freeze-out temperature, the
temperature at which they decouple from the plasma, and it is usually
assumed that this happens while the Universe is radiation
dominated. That is  the case when the reheating temperature
is much larger than the typical mass scale of the particle, that is,
reheating ended much before they freeze-out. 
However, as shown in Ref. \cite{tmax2}, the situation changes in an
scenario with a low reheating temperature, such that freeze-out takes
place when the Universe is still matter dominated. 
In fact it is shown that, for a B-ino like lightest neutralino, the
cosmological constraints on the B-ino mass and/or the right handed
slepton mass are relaxed and even disappear once a reheating
temperature below the B-ino mass is allowed. In our type of scenario,
moreover, the Higgses will be kept for a while out-of-equilibrium, due
to preheating, and therefore neutralinos will be produced also
out-of-equilibrium, initially most likely in a matter dominated
universe. If they do not re-enter equilibrium before reheating is
complete, their relic abundance may be different than that obtained  
in other scenarios.

Under the above assumptions, the evolution of the energy densities
(singlets, axions and radiation) 
and number densities (Higgses and neutralinos) during reheating can be
described by a simple 
set of equations \cite{kolb,tmax1,tmax2}: 
\begin{eqnarray}
\dot{\rho}_\phi&=& - 3 H \rho_\phi - \Gamma_{\phi a} \rho_\phi -
\Gamma_{preh} \rho_\phi  \,, \\
\dot{\rho}_{axion} &=& - 4 H \rho_{axion} + \Gamma_{\phi a} \rho_\phi \,, \\ 
\dot{n}_{H} &=& - 3 H n_H 
- (1- B_\chi) \langle \Gamma_H \rangle (n_H - n_H^{eq})  
- B_\chi \langle \Gamma_H \rangle (n_H -
\frac{n_\chi^2}{n_\chi^{eq^2}} n_H^{eq})  \nonumber \\
& & - \langle \sigma_H |v| \rangle (n_H^2 - n_H^{eq^2}) +
\Gamma_{preh} \frac{\rho_\phi}{M_H} \,, \\
\dot{n}_{\chi} &=& - 3 H n_\chi 
+B_\chi \langle \Gamma_H \rangle (n_H - \frac{n_\chi^2}{n_\chi^{eq^2}}
n_H^{eq})   
- \langle \sigma_\chi |v| \rangle (n_\chi^2 - n_\chi^{eq^2}) \,, \\
\dot{\rho}_{R} &=& -4 H \rho_{R} 
+(1 -B_\chi) \langle \Gamma_H \rangle \langle E_H \rangle (n_H - n_H^{eq})
\nonumber \\
& & + 2 \langle \sigma_H |v| \rangle \langle E_H \rangle (n_H^2 -
n_H^{eq^2}) 
+ 2 \langle \sigma_\chi |v| \rangle \langle E_\chi \rangle (n_\chi^2 -
n_\chi^{eq^2})  
 \,, 
\end{eqnarray}
where  $\Gamma_{preh}$ models the rate of
production of Higgses through ``inefficient'' preheating, $\langle
E_H \rangle$ and  $\langle E_\chi \rangle$ are the average energy per
Higgs and neutralino respectively, with  
\begin{equation}
\langle E_i \rangle \simeq \sqrt{m_i^2 + T^2} \;\;, i=H,\chi \,,
\end{equation}
and in the numerical calculations we have set the masses 
$m_H=m_\phi= 10 m_\chi = 1\, TeV$, and 
\begin{eqnarray}
\langle \Gamma_H \rangle \simeq \alpha_H m_H (1 - exp(-m_H/T)) \,, \\
\langle \sigma_i |v| \rangle \simeq \frac{ \alpha_i^2}{T^2} ( 1-
exp(-T^2/m_i^2)) \,,
\end{eqnarray}
with $\alpha_H=\alpha_\chi = 10^{-3}$. 
The preheating period would set the initial conditions to solve these
equations, and at $t_0=0$ we will take:
\begin{equation} 
\rho_\phi \simeq \rho_R \simeq \langle E_H \rangle n_H\simeq
\frac{V(0)}{3} \,.
\end{equation}
When reheating ends at $t=t_{RH}$, the temperature of the Universe
will be $T_{RH} \propto \sqrt{\Gamma_{\phi} M_P}$, where $M_P$ is the
reduced Planck mass and $\Gamma_\phi= \Gamma_{\phi a} + \Gamma_{preh}$
the total decay rate of the singlets. 

To start with, the Higgses will soon approach equilibrium due to
scattering process, with  $\langle \sigma_H |v| \rangle n_H^{eq} >
H(t_0)$. 
If $\Gamma_{preh} \ll 
\Gamma_\phi$, no appreciable amount of Higgses/radiation is further produced
and the Higgses will later decay in equilibrium.
This means that $\rho_R$ will be redshifted 
as $a^{-4}$, faster than the axions  which are produced in singlets
decay, $\rho_{axion} \propto a^{-3/2}$, and when
reheating ends we will have an axion-dominated Universe. 
Given that the entropy would be conserved, and that
$\rho_{axion}(t_{RH}) \simeq \rho_\phi(t_{RH})$, one can estimated the final
ratio of the energy density of axions to radiation as:
\begin{equation}
\frac{\rho_{axion}(T_{RH})}{\rho_R(T_{RH})} \simeq
\frac{\rho_\phi(t_0)}{\rho_R(t_0)} \left(\frac{T_0}{T_{RH}} \right)
\approx 10^{8} \label{fracaxion}\,.
\end{equation}
The relativistic axions will behave as an extra generation of
neutrinos at the time of 
nucleosynthesis, with the number of extra generations constraint by
$\delta N_\nu \le 1.8$ 
\cite{eta}. This translates into a bound for the energy
density of axions relative to that in radiation:
\begin{equation}
\left.  \frac{7}{8} \frac{\rho_{axion}}{\rho_R} \right |_{nucl.} \le
1.8 \,.
\label{bound}
\end{equation}
Comparing this bound with Eq. (\ref{fracaxion}), 
it is clear that we need the Higgses out of equilibrium and start decaying
into radiation at some time $t_1$ much before reheating is complete. This
will be achieved taking $\Gamma_{preh} > \Gamma_{\phi a}$.
This also means that allowing the standard
perturbative decay of the singlets into Higgses does not solve the
problem of axion dominance, since one expects the respective branching
ratios into axions and Higgses to be at most of the same order. We need some
extra effect apart from ``perturbative'' decay. 
However, if we take $\Gamma_{preh}$ as a constant parameter, and larger than
$\Gamma_{\phi a}$, we are forcing the singlets to decay completely
through preheating, which in a realistic scenario is unlikely to
happen. It is more  
reasonable to consider this parameter as a decreasing function of
time: as the energy density of the oscillating singlets decreases, the
rate of production through preheating will also diminish. We will
consider a simple ansatz where $\Gamma_{preh}$ follows an
exponential law, 
\begin{equation}
\Gamma_{preh}(t)= \Gamma_{preh}(t_0) e^{-b t} \,,
\end{equation}
with $b$  a constant such that $b < \Gamma_{preh}(t_0)^{-1}$. 
We can imagine this like  $\Gamma_{preh}(t)$ being switched on for a
while, until the energy density $\rho_\phi$ diminishes enough to make
parametric resonance completely negligible at $t > b^{-1}$, 
when $\Gamma_{preh}(t)$ is switched off.  

The situation now is as follows:  the Higgses  start decaying at a time
$t_1 < t_{RH}$, which can be estimated as:
\begin{equation}
t_1^{-1} \sim H(t_1) \sim \frac{\rho_\phi(t_1)}{n_H(t_1)}\frac
 {\Gamma_{preh}(t_0)}{M_H} \sim \frac{T_0}{M_H} \Gamma_{preh}(t_0) \,,
\end{equation}
and given that until this time the entropy is constant, and $T\propto
a^{-1}$, we have:  
\begin{equation}
T_1 \approx \left(.1 \frac{M_P}{M_H}\right)^{2/3} T_0^{1/3} 
\Gamma_{preh}^{2/3} \sim \left( 10^{20} \Gamma_{preh} \right)^{2/3}
\, GeV \label{t1}\,. 
\end{equation} 
Higgs/radiation  production stops at $t_2 \simeq
\Gamma_{preh}(t_0)^{-1}$. Inmediately after the Higgses
decay. Therefore, entropy is only released between $T_1$ and $T_2$, 
and in that interval the ratio $\rho_{axion}/\rho_R$
is frozen. After $t_2$ we are left still with the singlets producing
more axions, so the ratio of axions to radiation again increases until
$t_{RH}$ when is frozen again.

\begin{figure}[t]
\epsfxsize=10cm
\epsfxsize=10cm
\hfil \epsfbox{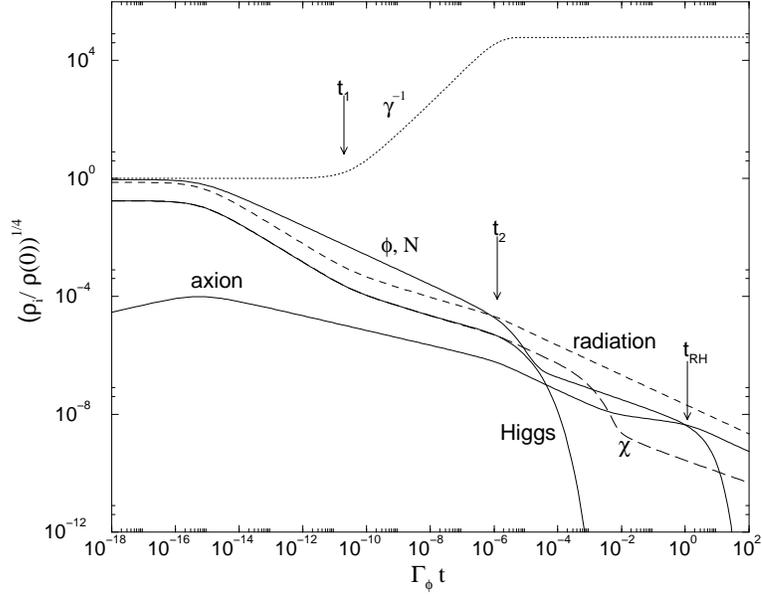} \hfil
\caption{{\footnotesize Evolution of the energy densities of the
singlets ($\phi,\,N$), axions, Higgses, radiation (dashed line) and
neutralinos (long-dashed line), when $\Gamma_{preh}(t_0)= 10^{-12} \,
GeV$, and $b = 10^{5} \Gamma_\phi  $. 
We have taken $B_\chi=0.5$. We have also included the factor
$\gamma^{-1} = S_f/S_i$ (dotted line).}}  
\label{fig1}
\end{figure}

This can be seen in Fig. \ref{fig1}, where we have plotted the
evolution of the energy densities for 
the case $\Gamma_{preh}(t_0)= 10^{-12} > \Gamma_{\phi}$  and
$b= 10^5 \Gamma_\phi$ as an
example. The time scale is given in units of $\Gamma_\phi$. The
singlets start decaying through $\Gamma_{preh}$ at $t\simeq t_1$, but
still we have assumed that they initially dominated the energy density
of the Universe. 
The Higgses initially are in equilibrium, but due to
the contribution from $\Gamma_{preh}$ they start to decay into
radiation and neutralinos. At this point the ratio $S_f/S_i$ starts to
increase as can be seen in the figure. Entropy production stops at
$t_2$, when the singlets energy density is partially depleted due to
$\Gamma_{preh}$. The Universe becomes radiation dominated at a
temperature $T_2 \simeq T_{RD} \simeq 10^{3}\, GeV$; this is larger
than  the lower reheating temperature we would expect only from
$\Gamma_{\phi a}$. 
Soon after, inefficient preheating is swichted
off and the Higgses decay. 
Until $\Gamma_\phi t \simeq O(1)$ we still produced more axions from
the remaining singlets. 
The  final ratio of the axions to radiation is $\rho_{axion}/\rho_R \sim
10^{-3}$, which is consistent with 
nucleosynthesis. The neutralinos follow the same evolution than the
Higgses as far as both are relativistic. Once the Higgses decay, they
go into equilibrium and when they become non-relativistic the ratio
$n_\chi/s$ freezes out. At $T_{RH}$ we have $(n_\chi/s)|_{RH} \simeq
4\times 10^{-11}$, which would imply a relic abundance of neutralinos
of order 1 today; they would dominate the dark matter in the Universe.
We notice that in this example the neutralinos enter into equilibrium
and freeze out in a radiation dominated Universe, so the calculation
of their relic abundance would not differ from the standard one, and
the usual bounds would apply.  

In the above example, the Universe becomes radiation dominated at a
temperature $T_2 < T_1$. However, from Eq. (\ref{t1}) we see that 
had we taken $\Gamma_{preh}(t_0) \geq 10^{-8}$, then $T_1 \sim T_0$, and
the Universe would be radiation dominated indeed during the whole
period of what we call ``reheating'', that to say, until the singlets
completely disappear.  

With this in mind,  we now turn to the calculation of the entropy
dilution factor $\gamma$: 
\begin{equation}
\gamma^{-1}= \left( \frac{S(T_{2})}{S(T_1)} \right) = 
\left( \frac{T_{2}}{T_{1}} \right)^{3} \left(
\frac{a(t_{2})}{a(t_1)} \right)^3 \,,
\end{equation}
which will depend on the initial value of $\Gamma_{preh}(t_0)$, but
not on its time dependence. 
If $\Gamma_{preh}(t_0) \leq 10^{-8}$, then the value of $T_2$ is given
by the condition $\rho_\phi(t_2) \simeq \rho_R(t_2)$, and we obtain:
\begin{equation}
\gamma^{-1} \approx 0.4 \left(
\frac{T_{2}}{T_{1}} \right)^3
\frac{\rho_\phi(t_1)}{\rho_R(t_{2})}\approx
\frac{T_0}{\sqrt{\Gamma_{preh(t_0)} M_P}} \leq 10^8\,.  
\end{equation}
The smaller the value of $\Gamma_{preh}(t_0)$, the larger the dilution
becomes.
On the other hand, for large values of $\Gamma_{preh}(t_0)$, entropy
production  takes place when the Universe is already
radiation dominated, the radiation energy density scales as $\propto
a^{-1}$, and the time $t_2$ is given by the condition $\rho_\phi(t_2)
\approx \rho_H(t_2)$ instead. In this case, $\gamma$ tends to a
constant value, given by:
\begin{equation}
\gamma^{-1}= \left(
\frac{a(t_{2})}{a(t_0)} \right)^3 \simeq \left(
\frac{\rho_\phi(t_0)}{\rho_H(t_0)}\right)^{3/2} \approx 10^4 
\times\left( \frac{\rho_\phi(t_0)}{\rho_R(t_0)}\right)^{3/2}
\,.
\end{equation}

Due to entropy production, the final ratio of axions to radiation
given in Eq. (\ref{fracaxion}) is also diluted, such that:
\begin{equation}
\frac{\rho_{axion}(T_{RH})}{\rho_R(T_{RH})} \simeq
\gamma \frac{\rho_\phi(t_0)}{\rho_R(t_0)} \left(\frac{T_0}{T_{RH}}
\right)\times exp(-F[\frac{\Gamma_{preh}(t_0)}{b}])\,.
\end{equation}
The last factor in the above equation is due to the
partial depletion in the energy density of the singlets around the
time $t \simeq \Gamma_{preh}(t_0)^{-1}$, with:
\begin{equation}
F[x]= x \left( 1 - exp(1/x) \right) \,.
\end{equation}

To summarize, in order to avoid axion dominance at the end of the
reheating period, we have allowed the singlets to decay into Higgses
through inefficient preheating, which will decay into radiation. 
The effect on the final ratio of axions to radiation is twofold: on one
hand it reduces the ratio because part of the singlets have been
converted into Higgses instead of axions; on the other hand, the final
ratio gets also diluted by a factor $\gamma$ due to entropy production.   
However, the same factor $\gamma$ will dilute the initial lepton
assymmetry produced in the s-neutrinos decays, which was our main
concern. The lower dilution factor is obtained when what we have
called ``reheating'' starts directly with a radiation dominated
Universe instead of the usual matter-inflaton dominated Universe. 
Based on simplest assumptions, we have obtained an upper
bound on $\gamma$, 
\begin{equation}
\gamma \sim 10^{-4} \times \frac{\rho_R(t_0)}{\rho_\phi(t_0)} \,,
\label{gamma} 
\end{equation}
which means that the baryon asymmetry at the end of reheating will be
at most:
\begin{equation}
 \frac{n_B}{s} \sim  10^{-12\pm 2} \,,
\end{equation}
a couple of orders of magnitude below the observational data. Given
that we have only considered a kind of toy model to study reheating, and the
uncertainties in it, we regard this result as quite promising.    
In particular, we remark again that the value of $\Gamma_{preh}$ is an
unknown in the model, controlled by the physics of
preheating. Moreover, we have taken $\rho_R(t_0) \simeq
\rho_\phi(t_0)$, but it may happen that we could produce more
radiation than expected before reheating starts.
Finally, we have considered a particular model for neutrinos masses as
a working example, such that the values of the
Yukawa couplings generate an asymmetry of the order $\epsilon_1 \sim
10^{-8}$.  
Other texture models with values of the Yukawas consistent
with the experimental data on neutrinos, and similar hierarchy among
the right handed neutrinos,  could give rise to a larger
asymmetry such as $\epsilon_1 \sim 10^{-6}$ \cite{example}, which
would give then the correct order of magnitude for the baryon
asymmetry.    

In any case, the less dilution we can have, the better, and this
translates into the Universe becoming radiation dominated as soon as
possible. In general, this means that  neutralino freeze out will take
place in a radiation dominated Universe, and the standard bound on
their relic abundance will apply.

Finally, we mention that we could also avoid axion dominance allowing
the singlets to decay into other light degrees of freedom apart from
axions. However,  
if the singlets decay into radiation this will imply a too large
dilution factor,  with \cite{kolb}  
\begin{equation}
 \gamma \sim  s_0 \frac{ \sqrt{\Gamma_\phi M_P}} { \rho_\phi (0)} \sim
10^{-8} \,,
\end{equation}
where the entropy density is $s_0 \sim
T_0^3$, and the  energy density in singlets $\rho_\phi$ are taken as the
 initial values at $t_0$.  

\section{Conclusion} 

In this paper we have discussed a realistic 
supersymmetric model of inflation \cite{us} which couples the inflaton 
to the Higgs, and when enlarged to include right-handed neutrinos allows
leptogenesis as the mechanism to generate the observed baryon
asymmetry of the Universe. From the particle physics point of view,    
the model is an extension of the Next-to-Minimal Supersymmetric Model
and  solves the $\mu$ problem via an intermediate scale which generates
the vev for the singlets. The interaction between the singlets and
matter fields are dictated by an approximate $U(1)$ Peccei-Quinn
symmetry, providing also a solution for the strong CP problem, and the
axions as a candidate for dark matter.  
As a hybrid inflationary model, it has a quite low scale for inflation
($O(10^8\,GeV$), and its predicts an spectral index $n=1$
consistent with the recent Boomerang and Maxima-1 data
\cite{boomerang}, and in principle a very low reheating temperature of
the order of a few GeV, barring the possibility of both GUT and
electroweak baryogenesis. Therefore we have extended the model
to include right-handed neutrinos, and have appealed to preheating
to produce the lightest right-handed sneutrino so that baryogenesis
may proceed via leptogenesis.

Extending the model to include heavy right handed neutrinos is one of the
preferred solutions not only for the sake of leptogenesis, but in
order to generate a light neutrino mass spectrum through the see-saw
mechanism, given the strong evidence from experiments in support of
such light masses and mixings. In this paper we have used as an example a
realistic model of all quark and lepton masses and mixing angles
\cite{steve1} based on single right handed
neutrino dominance \cite{steve}. 
We have chosen this particular example because it 
predicts not only a hierarchical spectrum
for the light neutrinos but also for the heavy right handed
neutrinos/sneutrino and in particular involves
a relatively light right handed state which is available for preheating, 
although we emphasise that this spectrum came
out of an analysis of neutrino masses and mixing angles which was not
performed with leptogenesis in mind.
The right handed neutrinos and sneutrinos
are too heavy to be produced thermally at
any stage in our inflationary model. However, we have argued that at least
the lightest right handed sneutrino can be produced through
parametric resonance during preheating, due to its couplings to the
Higgs doublets. Its CP-violating decay gives
rise to a lepton asymmetry, later converted by sphalerons into baryon
number. Note that the right handed sneutrinos do not couple directly
to the oscillating inflaton fields, but rather indirectly via the
Higgs doublets. Thus Higgs scalars are also expected to be produced
during reheating, and since these decay into radiation and
neutralinos we may estimate the amount of neutralinos and
entropy that is produced during preheating as shown in Figure \ref{fig1}.
Note that relativistic axions are also produced
during reheating but these are red-shifted away, although later on 
non-relativistic axions are additionally produced by the usual
misalignment mechanism and these will contribute to cold dark matter.

Because the heavy sneutrinos decay long before reheating is completed, 
the lepton asymmetry will be subsequently diluted by the entropy produced
in the decays of singlets and Higgses, before the time of
nucleosynthesis. In order to avoid too much dilution, we first
required the singlets to decay only into axions, which do not
thermalize and do not contribute to the radiation energy. The
radiation energy density has its origin in the out-of-equilibrium
decay of Higgses and sneutrinos, which have been previously produced
during preheating. If no more radiation is produced, by the time the
singlets completely decay the Universe becomes axion-dominated,
violating by many orders of the magnitude the bound on the number of extra
relativistic neutrino-like species at nucleosynthesis. We argue then
than ``inefficient'' preheating of the Higgses is required in order to
allow their out-of-equilibrium decay and some extra production of
radiation. Due to the long life-time of the singlets, it is possible
that, while they are oscillating, they could ``preheat''
other fields, at a rate similar or even larger than the perturbative
decay rate. To 
illustrate this point we have presented a simplified analyses of the
``reheating'' period, parametrising ``inefficient'' preheating by a
rate $\Gamma_{preh}$, in order to compare it with the perturbative
decay rate $\Gamma_\phi$, and it is the result of this
simplified analysis which is presented in Figure \ref{fig1}.
The main qualitative conclusion is that in order
to avoid axion dominance we would require $\Gamma_{preh} >
\Gamma_\phi$. Therefore, the main parameter controlling the analyses
is the ratio of these decay rates (the perturbative one and through
inefficient preheating), which will also depend to some extent on
the ratio of the Higgs and singlets masses. Note that the lighter
the Higgses are, the easier they will be produced. However, this may
modify the period of preheating following the end of
inflation. The more radiation which is produced the less the dilution we
will have later (Eq. (\ref{gamma}). 
 
To summarise, we have presented a semi-quantitative scenario for
leptogenesis in the context of a realistic supersymmetric
low scale hybrid inflationary model. A novelty of the model 
is that during the reheating period, the dilution of the
lepton-baryon asymmetry is not due to entropy produced in the
$inflaton$ decays but due to Higgs decay. This in turn is controlled by
how many Higgses we are able to ``preheat'' from the singlets before
reheating is finally completed, and it will help to avoid an
axion-dominated Universe at the end of reheating. 
In a model with a hierarchy in the
masses of the heavy right handed neutrinos, the combination of the
small asymmetry $\epsilon$ with some later dilution could give rise to
the correct order of magnitude for the final value of $n_B/s$. 
Note that in a realistic supersymmetric model such as this
the value of the baryon number is related to the question
of the nature and abundance of cold dark matter, and that these
questions are in turn related to the questions of 
the $\mu$ problem, the strong CP problem, and neutrino masses.

\begin{center}
{\bf Acknowledgements}
\end{center}
S.K. acknowledges the support of a PPARC Senior Fellowship.
We thank Vicente Di Clemente for reading the manuscript,
Toni Riotto for comments, and Gordy Kane for numerous comments and
encouragement.


\begin{thebibliography}{99}

\bibitem{yanagida1} M. Fukugita and T. Yanagida, Pjys. Lett. {\bf
B174} (1986) 45.  

\bibitem{luty} M. A. Luty, Phys. Rev. {\bf D45} (1992) 455.

\bibitem{yanagida2} H. Murayama, H. Suzuki and T. Yanagida,
Phys. Rev. Lett. {\bf 70} (1993) 1912. 

\bibitem{campbell} B. A. Campbell, S. Davidson and K. A. Olive, Nucl
Phys. {\bf B399} (1993) 111. 

\bibitem{others} H. Murayama and T. Yanagida, Phys. Lett. {\bf B322}
(1994) 349; T. Moroi and H. Murayama, hep-ph/9908223. 

\bibitem{kuzmin} G. t'Hooft, Phys. Rev. Lett. {\bf 37} (1976) 8;
V. A. Kuzmin, V. A. Rubakov and M. E. Shaposhnikov, 
Phys. Lett. {\bf B155} (1985) 36; J. Ambjrn, T. Askgaard, H. Porter
and M. E. Shaposhnikov, Nucl. Phys. {\bf B353} (1991) 346.  

\bibitem{superK} Super-Kamiokande collaboration, Phys. Lett. {\bf
B436} (1998) 33.; The Soudan 2 collaboration, Phys. Let.. {\bf B449}
(1999) 137; The MACRO collaboration, Phys. Lett. {\bf B434} (1998)
451. 

\bibitem{lepto} W. Buchm\"uller and T. Yanagida, Phys. Lett. {\bf
B302} (1993) 240;  W. Buchm\"uller and M. Pl\"umacher,
Phys. lett. {\bf B389} (1996) 73; M. Flanz, E. A. Paschos, U. Sarkar
and J. Weiss, Phys. Lett. {\bf B389} (1996) 693; A. Pilaftsis,
Phys. Rev. {\bf D56} (1997) 5431;  W. Buchm\"uller and M. Pl\"umacher,
Phys. Lett. {\bf B431} (1998) 354; W. Buchm\"uller and T. Yanagida,
Phys. Lett. {\bf B445} (1999) 399; J. Ellis, S. Lola and
D. V. Nanopoulos, hep-ph/9902364; M. S. Berger and B. Brahmachari,
hep-ph/9801271;  
R. Barbieri, P. Creminelli, A. Strumia and
N. Tetradis, hep-ph/9911315. 

\bibitem{gravitinos} For a review, see S. Sarkar,
Rep. Prog. Phys. {\bf 59} (1996) 1493; for the latest analysis, see
E. Holtmann, M. Kawasaki, K. Kohri and T. Moroi, Phys. Rev. {\bf D60}
(1999) 023506.  

\bibitem{kolb} E. W. Kolb and M. S. Turner, {\it The Early Universe},
(Addison-Wesley, Menlo Park, Ca. 1990). 

\bibitem{tmax1} D. J. Chung, E. W. Kolb and A. Riotto, Phys. Rev. {\bf
D60} (1999) 063504.

\bibitem{yanagida3} T. Asaka, K. Hamaguchi, M. Kawasaki and
T. Yanagida, hep-ph/9907559. 

\bibitem{peloso} G. F. Giudice, M. Peloso, A. Riotto and I. Tkachev,
JHEP {\bf 8} (1999) 14. 

\bibitem{us} M. Bastero-Gil and S. F. King,
Phys. Lett. \textbf{B 423} (1998) 27.

\bibitem{sugra} M. Bastero-Gil and S.F. King, Nucl. Phys. {\bf B549}
(1999) 391.

\bibitem{steve1}
S.F.King and M.Oliveira, hep-ph/0009287.

\bibitem{steve}
S.F.King, Phys. Lett. {\bf B439} (1998) 350;
S.F.King, Nucl. Phys. {\bf B562} (1999) 57;
S.F.King, Nucl. Phys. {\bf B576} (2000) 85;
S.F.King and N.N.Singh, Nucl. Phys. {\bf B591} (2000) 3.

\bibitem{covi} L. Covi, E. Roulet and F. Vissani, Phys. Lett. {\bf
B384} (1996) 169; L. Covi, E. Roulet and F. Vissani,
Phys. Lett. {\bf B424} (1998) 101.

\bibitem{epsilon} For review, see A. Pilaftsis, hep-ph/9812256;
W. Buchm\"uller and M. Pl\"umacher, hep-ph/9904310. 

\bibitem{example} See for example M. S. Berger and B. Brahmachari, in
\cite{lepto}. 

\bibitem{nbnbl} S. Yu. Khlebnikov and M. E. Shaposhnikov,
Nucl. Phys. {\bf B308} (1988) 885; J. S. Harvey and M. S. Turner,
Phys. Rev. {\bf D42} (1990) 3344. 

\bibitem{eta} See, for example, K. A. Olive, G. Steigman and
T. P. Walker, astro-ph/9905320. 

\bibitem{preheating1} J. H. Traschen, R. H. Branderberger,
Phys. Rev. \textbf{D 42} (1990) 2491; 
L. A. Kofman, A. D. Linde and A. A. Starobinsky,
Phys. Rev. Lett. \textbf{73} (1994) 3195; 
Y. Shtanov, J. Traschen and
R. Branderberger, Phys. Rev. \textbf{ D 51} (1995) 5438. 

\bibitem{preheating2} Lev Kofman, Andrei Linde, Alexei
Starobinsky, Phys. Rev. \textbf{D 56} (1997) 3258. 
S. Yu. Khlebnikov, I. I. Tkachev, 
Phys. Rev. Lett. \textbf{77} (1996) 219; 
T. Prokopec and T. G. Roos, Phys. Rev. {\bf D 55} (1997) 3768; 
S. Yu. Khlebnikov, I. I. Tkachev,
Phys. Rev. Lett. \textbf{79} (1997) 1607;
Juan Garc\'{\i}a-Bellido, Andrei Linde,
Phys.Rev. \textbf{D 57} (1998) 6075;

\bibitem{preheating3} J. Baacke, K. Heitmann and C. Patzold,
Phys. Rev. {\bf D58} (1998) 125013; 
G. Felder, L. Kofman and A. Linde,
hep-ph/9812289;  
P. B. Greene and L. Kofman, Phys. Lett {\bf B448} (1999); 
A. L. Maroto and A. Mazumdar, Phys. Rev. {\bf D59}. (1999) 083510. 

\bibitem{preheatingus} M. Bastero-Gil, S. F. King and J. Sanderson,
Phys. Rev. {\bf D60} (1999) 103517.  

\bibitem{gravitinos2} A. Maroto and A. Mazumdar, Phys. Rev. Lett. {\bf
84} (2000) 1655; R. Kallosh, L. Kofman, A. Linde and A. V. Proeyen,
Phys. Rev. {\bf D61} (2000) 103503; hep-th/0006354; 
G. F. Giudice, I. Tkachev and
A. Riotto, JHEP {\bf 08} (1999) 009; D. H. Lyth, Phys. Lett. {\bf
B469} (1999) 69; {\bf 476} (2000) 356; A. L. Maroto and J. R. Pelaez,
Phys. Rev. {\bf D62} (2000) 023518; 

\bibitem{BGM} M. Bastero-Gil and A. Mazumdar, hep-ph/0002004. 


\bibitem{ellis}  J. Ellis, K. Enqvist, D. V. Nanopoulos and
K. A. Olive, Phys. Lett. {\bf B191} (1987) 343. 

\bibitem{thermalaxions}
K. Choi, E. J. Chun and J. E. Kim, Phys. Lett. {\bf B403} (1997) 209. 

\bibitem{neutralinos} R. J. Scherrer and M. S. Turner, Phys. Rev. {\bf
D33} (1986) 1585; K. A. Olive and M. Srednicki, Phys. Lett. {\bf
B230} (1989) 78; R. Arnowitt and P. Nath, Phys. Rev. Lett. {\bf 70}
(1993) 3696; M. Drees and M. M. Nojiri, Phys. Rev. {\bf D47} (1993)
376;
G. L. Kane, C. Kolda, L. Roszkowski and J. D. Wells,
Phys. Rev. {\bf D49} (1994) 6173; J. Ellis, R. Falk and K. A. Olive,
Phys. Lett. {\bf B444} (1998) 367; J. Ellis, T. Falk, K. A. Olive and
M. Srednicki, Astropart. Phys. {\bf 13} (2000) 181.

\bibitem{tmax2} G. F. Giudice, E. W. Kolb and A. Riotto,
hep-ph/0005123. 

\bibitem{boomerang} P. de Bernardis et al. (Boomerang Coll.), Nature
{\bf 404} (2000) 955; S. Hanany et al. (Maxima Coll.), astro-ph/0005123

\end{thebibliography}
\end{document}